\documentclass[10pt,journal]{IEEEtran}
\ifCLASSOPTIONcompsoc
  \usepackage{cite}
\else
  \usepackage{cite}
\fi
\ifCLASSINFOpdf
  \usepackage[pdftex]{graphicx}
\else
\fi
\usepackage{amsmath,amssymb}

\usepackage{array}
\usepackage[caption=false,font=footnotesize]{subfig}
\usepackage{url}
\usepackage{booktabs}

\usepackage{hyperref}
\hypersetup{colorlinks,allcolors=black}
\usepackage{cleveref}

\usepackage{algorithm}
\usepackage{algpseudocode}

\mathchardef\simsym"0218

\algrenewcommand\textproc{}
\begin{document}
\title{Combined Generative and Predictive Modeling for Speech Super-resolution}
\author{Heming~Wang,~\IEEEmembership{Student Member,~IEEE},
        % Eric~Healy,~\IEEEmembership{Member,~IEEE}, and
        Eric~W. Healy, and
        DeLiang~Wang,~\IEEEmembership{Fellow,~IEEE}% <-this % stops a space
\IEEEcompsocitemizethanks{

\IEEEcompsocthanksitem H. Wang is with the Department of Computer Science and Engineering,
The Ohio State University, OH 43210, USA (e-mail: wang.11401@osu.edu)
\protect
% E-mail: 
\IEEEcompsocthanksitem E. W. Healy is with the Department of Speech and Hearing Science and the Center for Cognitive and Brain Sciences, 
The Ohio State University, Columbus, OH 43210, USA (e-mail: healy.66@osu.edu)
\protect
\IEEEcompsocthanksitem D. L. Wang is with the Department of Computer Science and Engineering and the Center for Cognitive and Brain Sciences, The Ohio State University, Columbus, OH 43210, USA (e-mail: dwang@cse.ohio-state.edu)}% <-this % stops an unwanted space
% \thanks{Manuscript received April 19, 2005; revised August 26, 2015.}}
}
\IEEEtitleabstractindextext{%
\begin{abstract}

Speech super-resolution (SR) is the task that restores high-resolution speech from low-resolution input. Existing models employ simulated data and constrained experimental settings, which limit generalization to real-world SR. Predictive models are known to perform well in fixed experimental settings, but can introduce artifacts in adverse conditions. On the other hand, generative models learn the distribution of target data and have a better capacity to perform well on unseen conditions.
In this study, we propose a novel two-stage approach that combines the strengths of predictive and generative models. Specifically, we employ a diffusion-based model that is conditioned on the output of a predictive model. Our experiments demonstrate that the model significantly outperforms single-stage counterparts and existing strong baselines on benchmark SR datasets. Furthermore, we introduce a repainting technique during the inference of the diffusion process, enabling the proposed model to regenerate high-frequency components even in mismatched conditions. 
An additional contribution is the collection of and evaluation on real SR recordings, using the same microphone at different native sampling rates. We make this dataset freely accessible, to accelerate progress towards real-world speech super-resolution.

\end{abstract}
}

\maketitle

% abstract/keywords data, the \IEEEtitleabstractindextext tex t will
\IEEEdisplaynontitleabstractindextext

% \fi
\IEEEpeerreviewmaketitle

\section{Introduction}
% \IEEEraisesectionheading{
% \section{Introduction}
% \label{sec:introduction}}


% for IEEE Computer Society journal papers produced under \LaTeX\ using
% IEEEtran.cls version 1.8b and later.

Speech super-resolution (SR), also known as speech bandwidth extension (BWE), aims to increase the sampling rate of low-resolution speech by generating high-frequency components.
It can improve the quality and intelligibility of the speech signal.
In addition, speech super-resolution has many potential applications in fields such as automatic speech recognition \cite{li2015dnn, albahri2016artificial}, hearing aids \cite{fullgrabe2010preliminary, van2020speech}, and text-to-speech synthesis \cite{nakamura2014mel}.
Speech SR has been addressed through various methods such as signal processing and deep learning methods.

Signal processing methods typically adopt the source-filter model \cite{milner2002speech}, which assumes that speech is produced by an excitation signal followed by a time-variant filter that simulates the vocal tract. The bandwidth extension is then divided into two separate tasks: spectral envelope extension and excitation signal extension.
Early studies employ codebook mapping techniques for spectral envelope extension, using two codebooks representing narrowband and wideband spectra \cite{unno2005robust, sadasivan2016joint}. Through training, the best matching entry in the wideband codebook to the narrowband codebook is used to generate the spectral envelope. Statistical methods such as the Gaussian Mixture Model (GMM) \cite{nour2008mel, nour2011memory} are also used. Unlike the codebook methods that rely on vector quantization and operate in a discrete space, GMM allows a continuous mapping of wideband coefficients from narrowband parameters. Hidden Markov Models (HMMs) \cite{song2009study, turan2015synchronous} are also used to estimate the spectral envelope, with each state representing the spectral envelope of a particular extension band. The probability density function of an HMM state is usually modeled by a GMM.
Excitation signal extension is an easier task. The most straightforward method is to perform spectral folding \cite{makhoul1979high, de2002yin} to extend the excitation signal from narrowband to wideband. In spectral folding, the missing upperband is generated by replicating the mirror image of the narrowband spectrum.
To preserve the harmonic structure of the excitation signal, one can employ sinusoidal synthesis \cite{chan1996wideband, abel2019sinusoidal} by utilizing a bank of oscillators in the narrowband. Sinusoidal synthesis is effective for excitation signals below 300 Hz. The excitation can also be enhanced by employing a nonlinear operation, which includes half-wave rectification, full-wave rectification \cite{iser2003neural}, and cubic spline \cite{mckinley1998cubic}.
Besides source-filter model-based methods are those regarding speech signals as the addition of consecutive amplitude and frequency modulation signals \cite{nagel2010modulation}, prediction of missing frequencies using error estimation, and non-negative matrix factorization \cite{sun2013nmf}. Additionally, alternative signal transforms like modified discrete cosine transform and the wavelet transform are explored for spectral analysis and extension \cite{park2011mdct,nizampatnam2017wavelet}.

Deep neural networks (DNNs) have been demonstrated to be superior to traditional statistical approaches for speech SR. Early DNN studies focus on speech magnitude, predicting wideband magnitude from narrowband magnitude using features such as log-power magnitudes \cite{li2015dnn, liu2015novel, gu2016speech}, vocal tract filter parameters \cite{botinhao2006frequency, kontio2007neural, pulakka2011bandwidth}, and cepstral coefficients \cite{abel2019sinusoidal}. The phase of the upperband is produced by flipping or copying the narrowband phase. Later studies address SR in the time domain \cite{gu2017waveform, kuleshov2017audio}, which naturally incorporates signal phase and yields better reconstruction quality. While these later studies achieve better objective metrics, the listening quality is closely related to that of magnitude estimation and does not improve significantly. Recent research has started to explore other signal representations.

For example, Lim et al. \cite{lim2018tf} proposed a time-frequency network (TFNet) that incorporates both time-domain and frequency-domain networks. These two networks are jointly optimized, and the output is fused to produce the final output. We proposed a time-domain convolutional network trained with a cross-domain loss function \cite{hm2021towards}. By leveraging cross-domain information, these networks are shown to outperform those based on a single representation domain. Lin et al. \cite{lin2021two} proposed a two-stage approach that combines the advantages of time- and frequency-domain methods. The first stage is a DNN that predicts the high-frequency content of a narrow-band input spectrogram, while the second stage refines the temporal details of the wide-band spectrogram obtained from the first stage.
% The proposed system achieves superior performance in speech enhancement and recognition compared to existing methods.

Another line of studies treats the SR task as a conditional generation problem. Rather than directly mapping LR (low-resolution) to HR (high-resolution) signals, these studies aim to learn a data distribution conditioned on LR features and then generate HR from the learned distribution.
% Specifically, speech synthesis techniques are employed to generate high-frequency components, for example, synthesizer-based methods by using the mel-spectrogram extracted from LR speech \cite{prenger2019waveglow, ling2018waveform, gupta2019speech, liu2022neural}.
Specifically, mel-spectrogram-based speech synthesizers are employed to generate HR speech by reconstructing mel-spectrograms from LR speech inputs \cite{prenger2019waveglow, ling2018waveform, gupta2019speech, liu2022neural}.
Recent studies have also adopted conditional generative adversarial networks (GANs) \cite{li2018speech, haws2019cyclegan, eskimez2019adversarial} or diffusion models that are conditioned on low-resolution signals \cite{lee2021nuwave, han2022nuwave2, yu2022diffsr} to extend the bandwidth.

% Challenges - 1 not one to one mapping 2 data is simulated
A major challenge in DNN-based super-resolution (SR) is training data acquisition. Most studies use simulation techniques to generate HR/LR pairs by downsampling HR signals. Although this approach is straightforward and can efficiently generate a large amount of data, the simulation process may introduce certain characteristics specific to the downsampling settings, limiting the robustness of the learned models in real-world application \cite{sulun2020filter,hm2021towards}.
Another challenge is rooted in the ill-posed nature of the SR problem, as the relationship of LR to HR signals is one-to-many, i.e. an LR signal may correspond to many valid HR signals. Predictive learning methods learn a one-to-one mapping from LR to HR, which can lead to overfitting to a specific simulation setup or produce overly smoothed speech spectrograms \cite{ling2015deep,hm2021towards}. This phenomenon is manifested in predictive models that perform well for certain types of downsampled signals but poorly for other downsampled signals, let alone challenging real-world scenarios.
Generative learning, learns a data distribution rather than a one-to-one mapping for the SR task, making it potentially more robust in realistic situations. However, its current SR performance is not competitive with mapping-based models.

In this study, we propose a two-stage approach to address the challenges of DNN-based SR.
Specifically, the first stage employs predictive learning to generate coarsely enhanced speech, which is fed as the conditioner to the second stage that employs a diffusion learning model.
Our proposed approach combines the advantages of both types of learning and achieves better SR performance with robustness to recorded data.
We also exploit a repainting technique that is widely used in diffusion-based learning, which improves model generalization to different simulation methods.
In addition, we record and publicize speech corpora by employing the same microphone at multiple native sampling rates.

The contributions of this study are three-fold. First, we propose a novel DNN architecture for speech SR that integrates predictive learning and generative learning. Second, the proposed approach is demonstrated to achieve better SR performance and robustness than other strong baselines. Third, we record and publicize multi-resolution datasets to facilitate future speech SR research in the community.

The rest of the paper is organized as follows.
Section \ref{sec:background} describes the background of diffusion models and the SR task.
Section \ref{sec:methods} presents the proposed approach.
Section \ref{sec:setup} describes the experimental setup and multi-resolution data recording. Evaluation results and comparisons are presented in Section \ref{sec:results}. Finally, concluding remarks are given in Section \ref{sec:conclusion}.

\section{Background}
\label{sec:background}

\subsection{Denoising Diffusion Probabilistic Model}

\begin{figure*}[!thb]
\centering
    \includegraphics[width=0.8\textwidth]{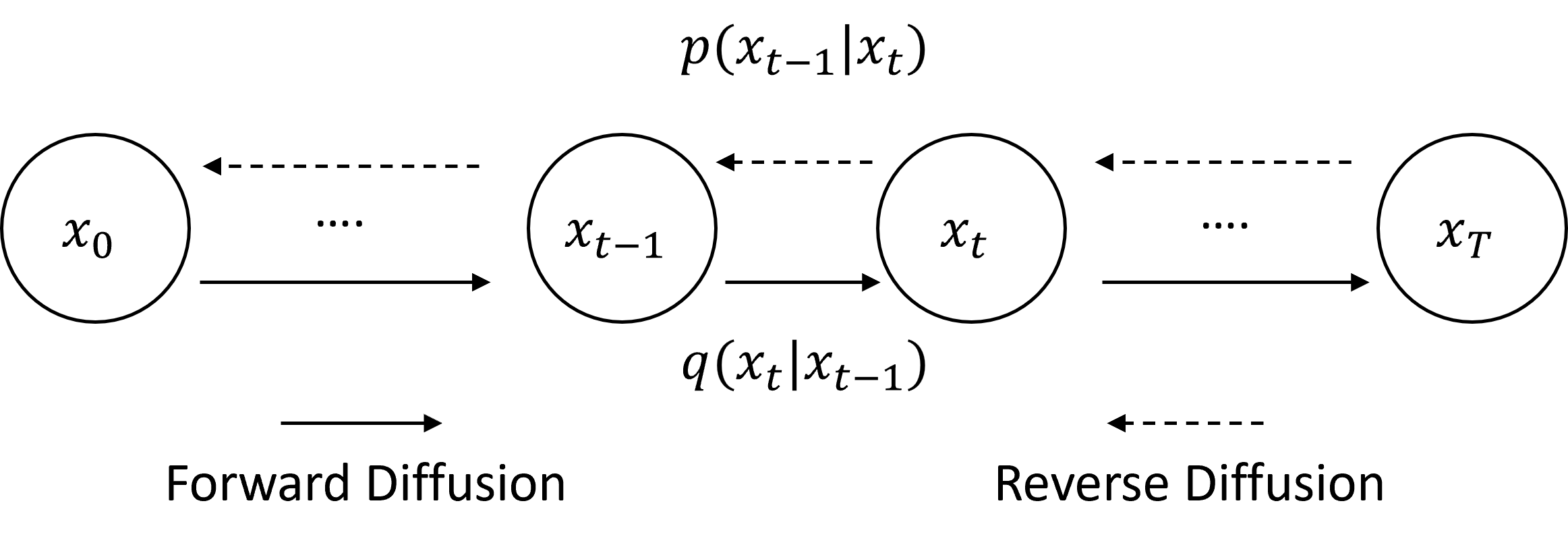}
    \caption{Illustration of the diffusion processes in a denoising diffusion probabilistic model.}
    \label{fig:diffusion}
\end{figure*}

A standard denoising diffusion probabilistic model (DDPM) \cite{ho2020denoising}, consists of a forward process and a reverse process, as depicted in Fig. \ref{fig:diffusion}.
The DDPM with $T$ time steps contains latent variables $x_0, x_1, ..., x_T$, all with the same dimensionality.
\subsubsection{Forward Process}
During the forward process, the joint distribution is defined as a Markov chain from $x_0$ to $x_T$,
\begin{equation}
    q(x_1,...,x_T|x_0) = \prod_{t=1}^T q(x_t|x_{t-1}).
\end{equation}
At each step a Gaussian noise is added, and the noise is controlled by the noise schedule parameter $\sigma(t)$ and $\gamma$, where
\begin{equation}
\sigma(t)^2 = \frac{\sigma^2_{min} (( \sigma_{max} / \sigma_{min})^{2t} - e^{-2\gamma t}) log(\sigma_{max} / \sigma_{min}) }{\gamma + log(\sigma_{max} / \sigma_{min})}
\end{equation}
is a fixed noise schedule that manipulates the Gaussian distribution in the forward process. A Gaussian noise is added to the previous sample $x_{t-1}$ during each step, gradually converting $x_0$ to an isotropic Gaussian distribution $p(x_T) = \mathcal{N}(0,I)$. Since each transition step follows a Gaussian distribution, $x_t$ can be directly computed from the distribution of $x_0$ by marginalizing $x_1, ..., x_{t-1}$, i.e.,
\begin{equation}
    q(x_t|x_0) = \mathcal{N} (x_t; \mu(x_0, y, t), \sigma(t)^2).
    \label{eqn:x0}
\end{equation}
As derived in \cite{richter2022sgmse}, the mean $\mu$ has a closed form solution,
\begin{align}
    \mu(x_0, y, t) = e^{-\gamma t} x_0 + (1-e^{-\gamma t}) y.
    \label{eqn:mutheta}
\end{align}

\subsubsection{Reverse Process}

The reverse process in contrast is to progressively remove the added noise from the latent variable $x_T$ to obtain $x_0$.
In the standard DDPM, this is done by subtracting an estimated noise from a DNN model. During training, the model is trained to predict the added noise given the diffused signal and the noise embedding information.
Other methods, such as the score-based diffusion, model the forward process using a stochastic differential equation (SDE). Then a DNN model is used to estimate the score, which is the gradient of the log probability density.
Numerical differential equation solvers are utilized to solve a reverse-time SDE using the estimated scores.
Repeating these steps gradually derives the target data $x_0$ from the whitened Gaussian noise $x_T$.

\subsubsection{Conditional Reverse Process}
Inspired by \cite{kawar2022ddrm,salimans2022progressive,croitoru2023diffusionsurvey}, instead of estimating noises or scores, we employ a DNN model to directly estimate the target data $x_{0t}$ at each time step $t$.
This approach has two major advantages. First, unlike the standard DDPM which requires either $\mathcal{L}_1$ or $\mathcal{L}_2$ norm on the noise estimation, we can employ an established training objective in the target domain and directly compute a loss on the target data $x_0$. Second, after the training is done, we have a model that effectively estimates $\hat{x}_{0t}$ given the time step $t$, which achieves faster inference compared with the standard DDPM process.

\subsection{Speech SR Problem Formulation}
\label{sec:formulation}
% Describe the super-resolution formulation.
Given a low-resolution speech segment $s^{lr}$ at a sampling rate of $\mathit{fs}^{lr}$, the objective of speech super-resolution is to reconstruct a high-resolution speech signal $s^{hr}$ at a higher sampling rate of $\mathit{fs}^{hr}$ such that $\mathit{fs}^{hr} > \mathit{fs}^{lr}$, thereby restoring high-frequency components.
The number $\mathit{fs}^{hr} / \mathit{fs}^{lr}$ is denoted as the \textit{upsampling ratio}.
To reconstruct the high-resolution signal, we first employ a predictive learning model $f$ that takes the upsampled low-resolution signal $s^{inp}$ as input, which is obtained by applying interpolation on the LR signal $s^{lr}$. With the model parameters denoted as $\phi$, the model produces the corresponding reconstruction $s^{pred}$: 
\begin{equation}
    s^{pred} = f(\phi, s^{inp}),
\end{equation} 
where ${s}^{pred}$ is the estimated SR signal obtained from the first stage.
The predicted SR output is then utilized along with the original LR input as the conditioner for the second stage of diffusion-based learning.
We denote the parameters of the diffusion-based model as $\theta$, and the reverse diffusion process as $g$. During the inference stage, the diffusion model aims to reconstruct the final output from the whitened Gaussian noises given the conditioners, which is formulated as
\begin{equation}
    \hat{s} = g(\theta, s^{pred}, s^{inp}).
\end{equation}

\section{Model Description}
\label{sec:methods}
\begin{figure*}[!thb]
\centering
    \includegraphics[width=1.0\textwidth]{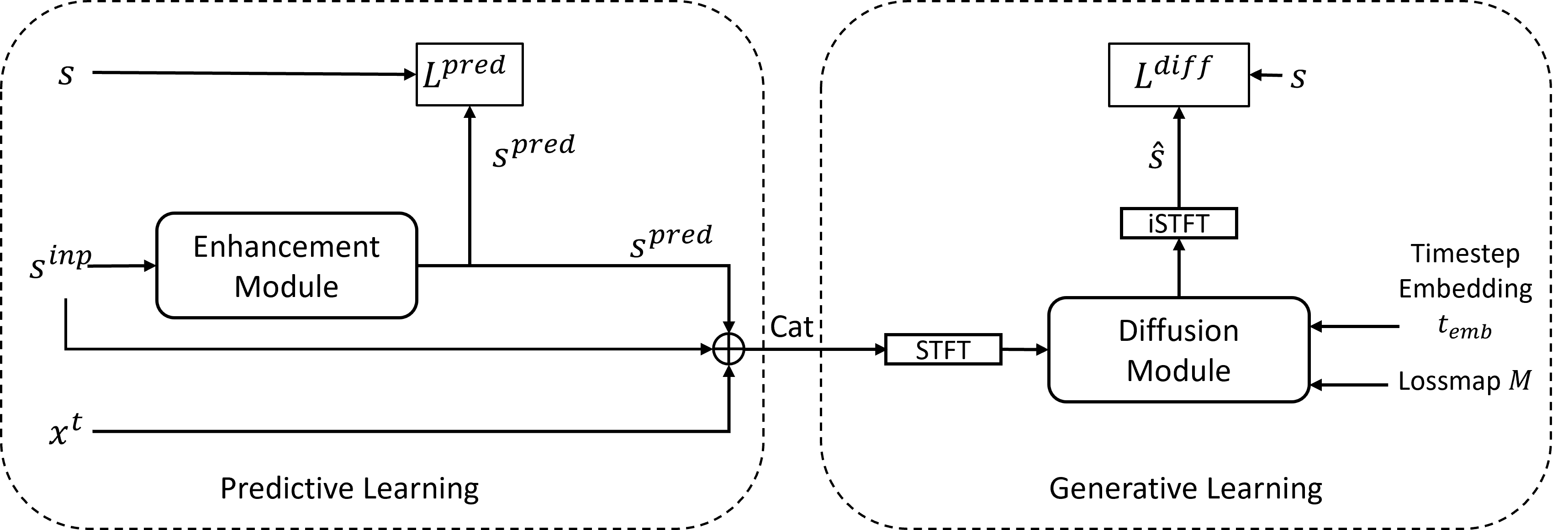}
    \caption{Two-stage model diagram that depicts the training procedure of the predictive learning stage and the generative learning stage.}
    \label{fig:pipeline}
\end{figure*}

% Introduce the methods used, two stage
We introduce a two-stage DNN model for speech super-resolution, as shown in Fig. \ref{fig:pipeline}.
In the first stage, we utilize a time-domain dual-path attentional recurrent network (DPARN) \cite{pandey2020dparn} for predictive learning. Its output conditions the second-stage diffusion learning.
For generative learning, we propose an attention-based residual convolutional network (ARCN) and estimate the added noise during the diffusion process. During inference, we use the enhanced speech from the first stage as the conditioner and progressively generate the missing bandwidth through the reverse diffusion process.
% The speech signals are chunked with segments of fixed length with overlap before feeding them to the diffusion module. 
Detailed descriptions are given in the following subsections.

\subsection{DNN Architectures}

\begin{figure*}[!thb]
\centering
\includegraphics[width=0.9\textwidth]{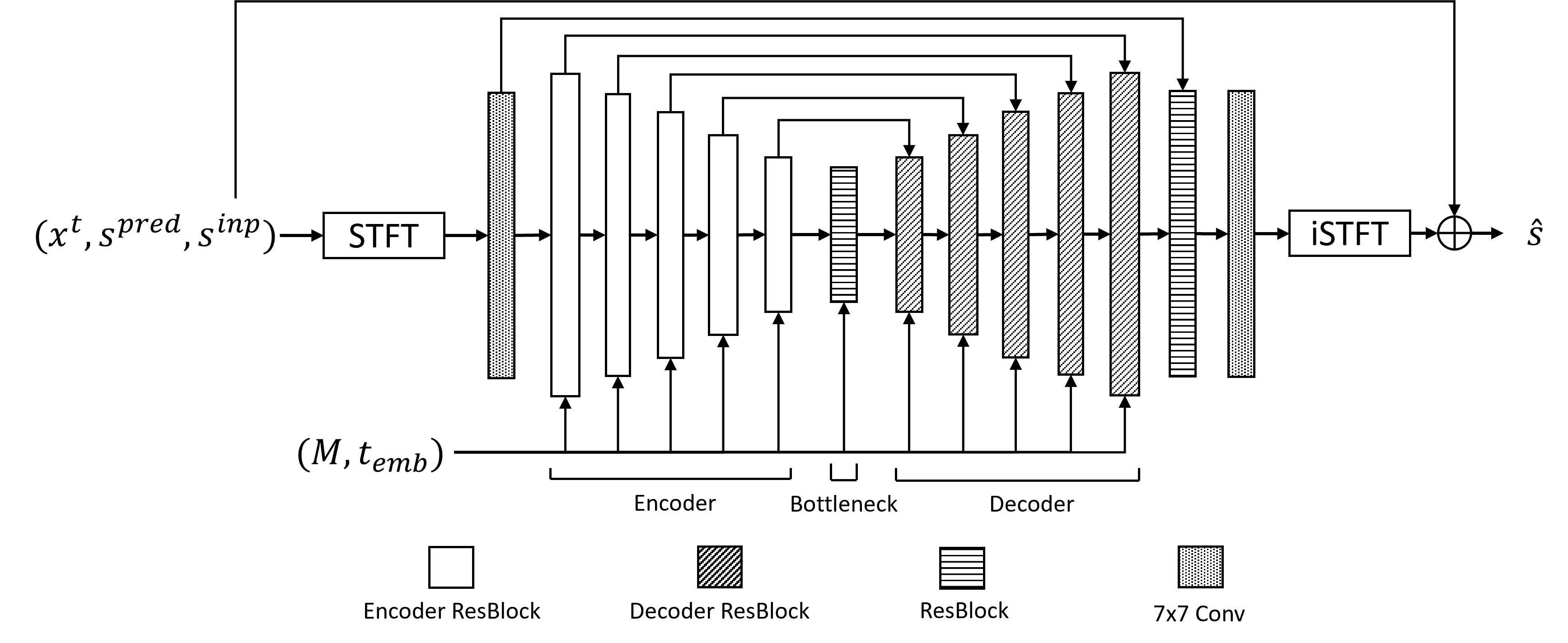}
\caption{Diagram of the proposed attentive residual convolutional network (ARCN) as the diffusion module, and ``ResBlock" denotes an attentional residual block.}
\label{fig:arcn}
\end{figure*}

\subsubsection{Predictive Learning}

For predictive learning, we employ the dual-path attentional recurrent neural network (DPARN) \cite{pandey2020dparn} as the enhancement module. DPARN is an enhanced version of the dual-path recurrent neural network, which was originally introduced for time-domain speaker separation \cite{luo2020dprnn}. In a dual-path network, utterances are divided into overlapping chunks and processed sequentially using intra-chunk and inter-chunk recurrent neural networks (RNNs) to efficiently handle the given time series. This approach reduces the sequence length for RNN modeling, improving training efficiency. Furthermore, it allows for a relatively small frame shift in time-domain speech processing, resulting in a significant performance gain. DPARN further incorporates inter-chunk and intra-chunk attention, resulting in notable improvements in performance and training efficiency. To reduce the computational burden, we make two modifications: i) adopting residual connections instead of dense connections between RNN modules and ii) limiting the number of DPARN blocks to two. More details of DPARN design can be found in \cite{pandey2020dparn}.

\subsubsection{Diffusion Module}
We implement the diffusion module using an ARCN, which is based on UNet. 
The UNet architecture has been extended in recent years to incorporate attention and residual blocks in diffusion-related models \cite{ho2020denoising,richter2022sgmse}.
However, directly adopting the standard UNet is unsuitable for audio signals, as it is designed to process the input of fixed dimensions.
ARCN enables variable input lengths and easier conditioner incorporation.

For diffusion training, ARCN requires five inputs: the diffused signal $x_{t}$, the enhanced speech signal $s^{pred}$ from the enhancement module, the LR speech input $s^{inp}$, a lossmap $M$, and the time step embedding $t_{emb}$ (see Fig. \ref{fig:arcn}). 
Prior to feeding these inputs to ARCN, we apply Short-Time Fourier Transform (STFT) to $x_{t}$, $s^{pred}$, and $s^{inp}$, concatenating their real and imaginary parts as separate convolutional channels.
The time step embedding and lossmap serve as local conditioners for the residual block. The time step embedding is generated using Fourier embedding in \cite{vaswani2017attention} followed by two linear layers. The lossmap $M$ consists of binary indicators (0 and 1), where time-frequency (T-F) units marked as ones are those to be super-resolved.

Our ARCN adopts an encoder-decoder architecture with a bottleneck block, as illustrated in Fig. \ref{fig:arcn}.
The encoder and decoder have symmetric designs, incorporating skip connections to enhance feature reusability and combat the vanishing gradient issue. 
% Specifically, we concatenate each encoder block output with the input from the corresponding decoder block.
Instead of standard convolution layers, we build the encoder, decoder and bottleneck with attentional residual blocks to improve learning of temporal dependencies between frames.
Residual connections within each block facilitate feature reuse and accelerate training convergence. 
Operating on complex spectrogram vectors obtained from STFT, the model concatenates real and imaginary parts into a 3-dimensional representation ($C \times T \times F$), where $C$ represents the number of convolutional channels (6 channels in this study), $T$ the number of time frames and $F$ the size of the feature dimension. 
The input is initially transformed into 64 convolutional channels using the 7x7 input convolution layer.
The 64-channel input undergoes processing through 5 attentional residual blocks in the encoder, 1 bottleneck block, and then 5 attentional residual blocks in the decoder.
The decoder output undergoes another residual block, and is then processed with a 7x7 convolutional layer.
Finally, the speech SR estimate, $\hat{s}$, is derived by adding the LR input $s^{inp}$ with the inverse STFT (iSTFT) applied to the output.

\begin{figure}[!thb]
\centering
\includegraphics[width=0.99\linewidth]{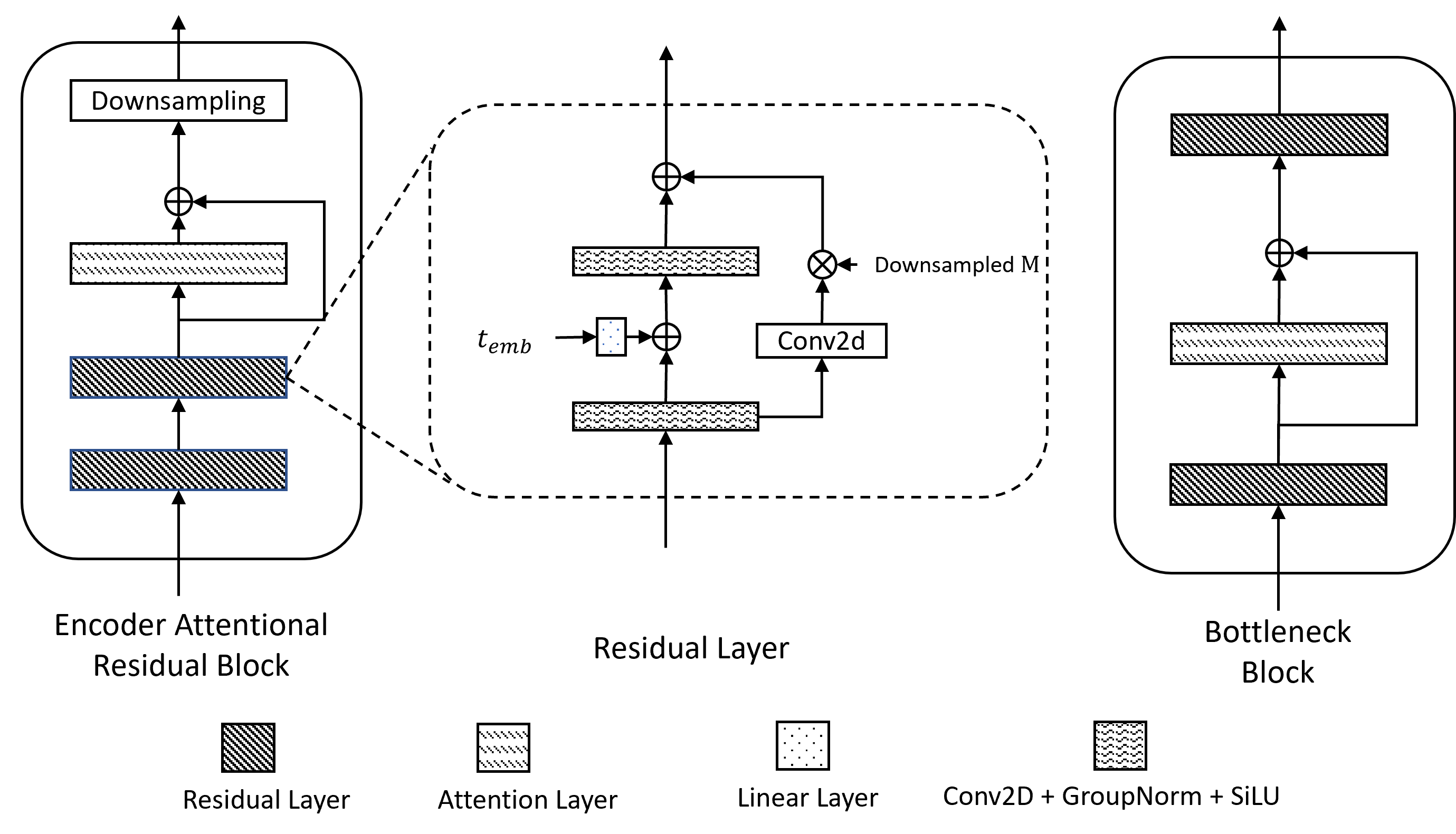}
\caption{Diagrams showing the detailed design of an attentional residual block within the ARCN encoder.}
\label{fig:residual}
\end{figure}

% attention residual block
Fig. \ref{fig:residual} shows the details of an attention residual block in the encoder. It comprises two residual layers, a lightweight attention layer, and a downsampling operation.
The output of the attention layer is added to its input using a residual connection.
To prevent the checkerboard artifact \cite{odena2016checkerboard}, we implement the downsampling operation using finite impulse response filters \cite{zhang2019fir}. 
Each residual layer in our model comprises two convolution layers, and utilizes the time step embedding and lossmap as local conditioners.
Each convolution layer consists of a 2-dimensional (2-D) convolution, group normalization \cite{wu2018group} and Sigmoid Linear Unit (SiLU) nonlinearity \cite{elfwing2018silu}.
A convolution layer has a kernel size of 1x3 (time x frequency), zero-padding of size 1 applied to each side of the feature maps along the frequency dimension, and 64 output channels.
% The time step embedding, denoted as $t_{emb}$, is obtained using the Fourier embedding approach \cite{vaswani2017attention}. This approach involves projecting the time step $t$ to a higher-dimensional vector through a linear layer and adding it as a bias term after the first convolution layer.
The downsampled lossmap is incorporated through its multiplication with the output of the first convolution layer after a pointwise convolution.
A residual block in the decoder is similar, but uses an upsampling operation instead of downsampling. 
As illustrated in Fig. \ref{fig:residual}, the bottleneck block has a different arrangement of residual layers.

\begin{figure}[!thb]
\centering
\includegraphics[width=0.99\linewidth]{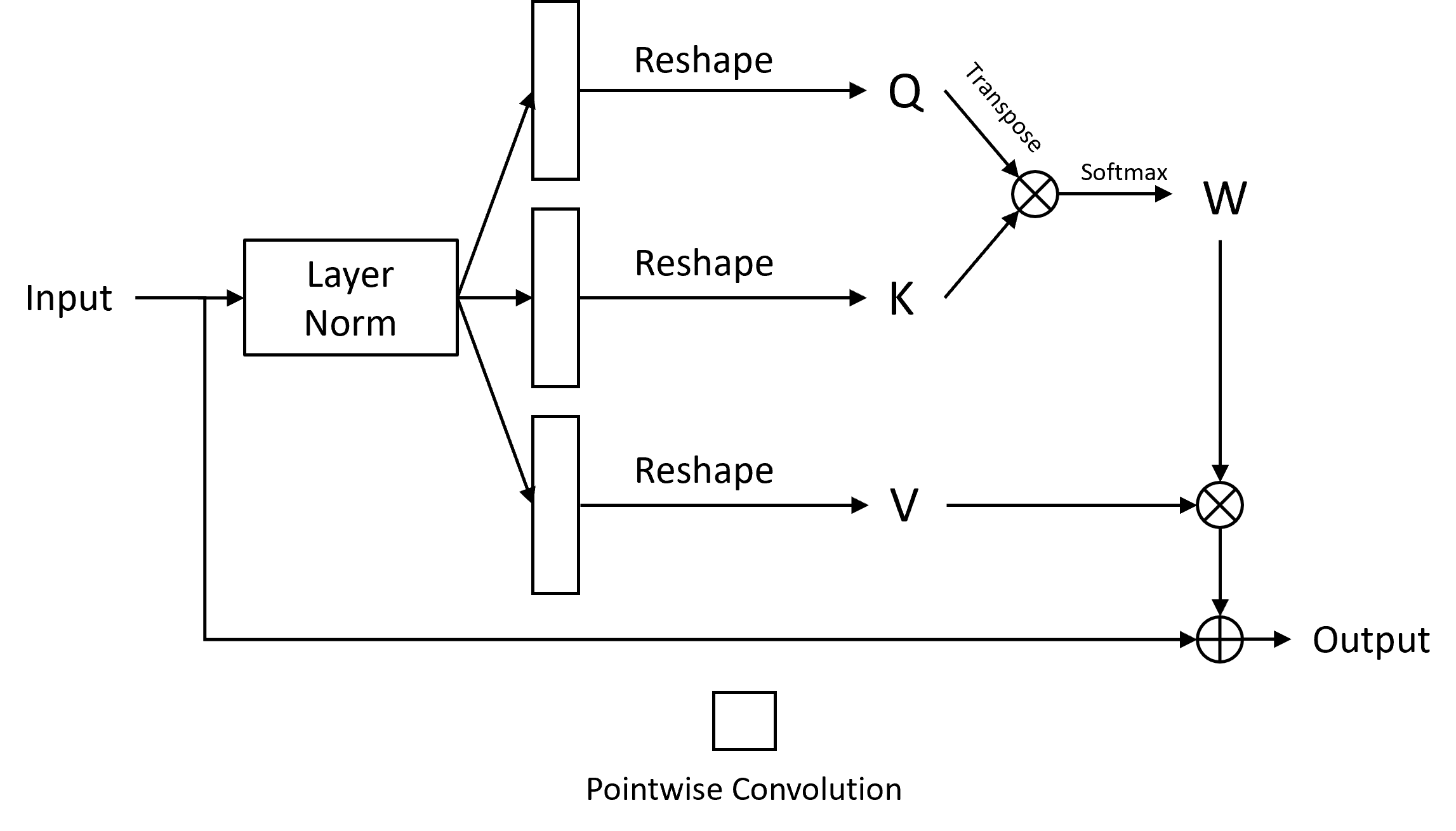}
\caption{The architecture of an attention layer.}
\label{fig:attn}
\end{figure}

% 
% Attention block
Fig. \ref{fig:attn} shows the architecture of an attention layer, which operates on an input of shape $C \times T \times F$.
The attention layer transforms the input using three pointwise convolution layers to derive the query ($\textbf{Q}$), key ($\textbf{K}$), and value ($\textbf{V}$) tensors of $E \times T \times F$, $E \times T \times F$, and $C \times T \times F$ respectively, where $E$ is set to 5 to reduce the computational burden.
They are subsequently rearranged into 2-D matrices $T \times E \cdot F$, $T \times E \cdot F$, and $T \times C \cdot F$, respectively.
The correlation scores between the query and key matrices are calculated through matrix multiplication $\textbf{W} = \textbf{Q}\textbf{K}^T$, where superscript $T$ denotes matrix transpose. This produces $\textbf{W}$ with dimensions $T \times T$.
The attention score is then computed by applying a Softmax function to $\textbf{W}$, and multiplied with the value matrix $\textbf{V}$:
\begin{equation}
\textbf{A} = \text{Softmax}(\textbf{W}) \textbf{V}.
\end{equation}
The result, denoted as the attentive output ($\textbf{A}$), emphasizes important time frames. This mechanism enables ARCN to capture long-term temporal dependencies.
To preserve information from the original input, a residual connection is employed to merge the attentive output with the input as shown in Fig. \ref{fig:attn}.

\subsection{Loss Functions}
We simultaneously optimize two modules through joint training. The predictive learning module utilizes a complex-domain loss (denoted as $\mathcal{L}^{pred}$) proposed in \cite{zq2020complex} to predict SR speech. 
This loss function is defined in terms of  $\mathcal{L}_{1}$ differences in magnitude, and real and imaginary spectrograms:
\begin{align}
\mathcal{L}^{pred} = \frac{1}{TF} \sum_{t=1}^T \sum_{f=1}^F & \big[ ||S^{pred}(t,f)| - |S(t,f)|| + \nonumber \\
(|S^{pred}_r(t, f) - S_r(t, f)| &+ |S^{pred}_i(t, f) - S_i(t, f)|) \big].
\end{align}
Here, $T$ and $F$ denote the total number of time frames and frequency bins, indexed by $t$ and $f$, respectively. $S^{pred}$ and $S$ represent the STFTs of the predicted and original speech signals, respectively. Subscripts $r$ and $i$ refer to the real and imaginary parts of the complex vectors, respectively, and $|\cdot|$ measures the magnitude. To compute the STFT vectors $S^{pred}$ and $S$, the waveforms are first divided into 32-ms frames with a frame shift of 8 ms and then multiplied by a Hanning window.

For the diffusion module, we adopt a T-F loss proposed in \cite{wang2020tfloss}:
\begin{align}
    \mathcal{L}^{T}(\hat{s}, s) &= \frac{1}{N} \sum_{n=1}^N | \hat{s}(n) - s(n) |  \nonumber \\
    \mathcal{L}^{F}(\hat{S}, S) &= \frac{1}{TF} \sum_{t=1}^T \sum_{f=1}^F ||\hat{S}(t,f)| - |S(t, f)| |  \nonumber \\
    \mathcal{L}^{diff} &= \alpha \mathcal{L}^{T} + (1-\alpha) \mathcal{L}^{F}.
\end{align}
The time domain loss $\mathcal{L}^{T}$ is the $\mathcal{L}_1$ difference between SR and HR signals ($\hat{s}$ and $s$), where $n$ to indexes time samples.
The frequency domain loss $\mathcal{L}^{F}$ compares the magnitudes of the STFTs of SR and HR signals.
The T-F loss combines $\mathcal{L}^{T}$ and $\mathcal{L}^{F}$, and we empirically set $\alpha$ to 0.85.

The overall loss for joint training of the two modules is given by:
\begin{equation}
\mathcal{L} = \mathcal{L}^{pred} + \lambda(\tau) \mathcal{L}^{diff}.
\end{equation}
We use a time varying $\lambda(\tau) = 1 / (e^\tau - 1)$ to control the weight of the diffusion loss $\mathcal{L}^{diff}$ at different diffusion steps $\tau$ \cite{lu2023understanding}.

\subsection{Training and Inference}

\begin{algorithm}
\caption{\hspace{-1.3mm}. Training Procedure}
\label{alg:train}
\begin{algorithmic}
\State $s^{inp} = Upsample(Downsample(Filtering(s^{hr})))$
\State $s^{pred} = f(\phi, s^{inp})$
\State Sample $t \sim Uniform({1,2,...,T})$
\State Sample $z \sim \mathcal{N}(0,I)$
\State $x_t \leftarrow \mu (s^{hr}, s^{inp}, t) + \sigma(t)^2 z $
% \State $\hat{x}_{0t} \leftarrow g(\theta, x_t, s^{pred}, s^{inp}, t)$ 
\State $\hat{s} \leftarrow g(\theta, x_t, s^{pred}, s^{inp}, t)$ 
\State Take gradient descent on
\State $\mathcal{L}^{pred} + \lambda(t) \mathcal{L}^{diff}$
\end{algorithmic}
\end{algorithm}
The training procedure of the proposed approach is summarized in Algorithm \ref{alg:train}.
The coarsely enhanced speech $s^{pred}$ is employed as the conditioner for the diffusion model.
For diffusion training, we randomly sample a time step $t$, and use it to generate the time step embedding.
The model $g$ will estimate the target SR speech at the given diffusion step.
Gradient descent serves to optimize the predictive model $f$ and the generative model $g$ jointly.

\algrenewcommand\textproc{}

\begin{algorithm}
\caption{\hspace{-1.2mm}. Inference Procedure}
\label{alg:infer}
\begin{algorithmic}
\Procedure{Resample}{$x$}
  \State Return Upsample(Downsample(Filtering($x$)))
\EndProcedure
\State $s^{inp} \leftarrow Upsample(s^{lr})$
\State $s^{pred} \leftarrow f(\phi, s^{inp})$
\State Initialize $ x_{0T} \leftarrow s^{pred}$

\For{$t = T-1,T-2,... ,0$}
    \State Sample $z \sim \mathcal{N}(0,I)$
    \State $ x_{t} \leftarrow \mu(x_{0t+1}, s^{inp}, t) + \sigma(t)^2 z$
    \State $ x_{0t} \leftarrow g(\theta, x_{t}, s^{pred}, s^{inp})$
    \State $ x_{0t} \leftarrow s^{inp} + (x_{0t} - Resample(x_{0t}))$ \Comment{Repainting}
\EndFor
\State \Return $\hat{s} \leftarrow x_{00}$
\end{algorithmic}
\end{algorithm}

Algorithm \ref{alg:infer} outlines the inference process, featuring two key modifications from the standard DDPM reverse process.
First, we initialize the diffused input signal $x_{0T}$ using the predicted speech $s^{pred}$ obtained in the first stage, instead of white Gaussian noise.
Second, we leverage the low-frequency components of $s^{inp}$ to guide the reverse diffusion process. 
Utilizing given narrowband components, we inject LR information into the sampling process by replacing the low-frequency region with the ground truth.
Specifically, we apply the same set of downsampling and upsampling operations employed during training (the $Resample$ operation defined in the algorithm), i.e.,
\begin{equation}
     x_{0t} = s^{inp} + (x_{0t} - Resample(x_{0t})).
\end{equation}
Here we essentially combine the high-frequency part of the SR prediction $x_{0t}$ at time step $t$, with the low-frequency part of the input signal $s^{inp}$.
This process is similar to the repainting technique in the image domain \cite{lugmayr2022repaint}, and applied in DiffWave based SR \cite{yu2022diffsr}.
Different from repainting, our approach operates in the time domain. Compared to \cite{yu2022diffsr}, we do not subtract a gradient term in each reverse diffusion step.
In addition, we initialize the reverse diffusion process with the output of the prediction module $s^{pred}$, which is referred to as the shallow diffusion mechanism in \cite{liu2022diffsinger}.
Our inference algorithm not only yields improved objective scores but also expedites the diffusion process.

\section{Experimental Setup}
\label{sec:setup}
\subsection{Simulated Dataset}
\label{ssec:data}
We conduct our experiments on the VCTK dataset \cite{VCTK2017cstr}, which contains 44 hours of speech recordings from 108 speakers, and follow the task design in \cite{han2022nuwave2}.
% Only the \textit{mic1} recordings of the VCTK dataset are used for experiments and recordings of p280 and p315 are excluded. We train the model on the first 100 speakers and test it on the remaining 8 speakers.
We exclusively utilize the \textit{mic1} recordings from the VCTK dataset for our experiments, while omitting the recordings of speakers p280 and p315. Our model is trained on the first 98 speakers, validated on the excluded two speakers, and tested on the remaining 8 speakers.
We preprocess the utterances as follows.
First, we resample them to 16 kHz if they have a higher sampling rate. Second, we normalize them to have zero mean and unit variance; as suggested in \cite{hm2021towards} this helps with generalization to untrained data.
To generate LR speech, we first convolve the HR speech with an eighth-order Chebyshev type I lowpass filter, and then subsample it to a desired sampling frequency.
We then use cubic spline interpolation to upsample generated LR signals to match the lengths of the corresponding HR signals before feeding them to the proposed model.

\subsection{Recorded Datasets}
\label{sec:real_dataset}
Although experiments on simulated data show superior performance for DNN-based models, speech SR performance in actual applications may be significantly worse due to the mismatch between synthetic and recorded data.
In a previous study \cite{hm2021towards}, we found that trained SR models are very sensitive to recording channels and downsampling schemes.

To overcome the limitations of simulated data, we created recordings of two datasets, Device and Produced Speech (DAPS) \cite{mysore2014daps} and VCTK \cite{VCTK2017cstr}, each at different sampling rates.
The DAPS dataset comprises carefully aligned speech recordings, including studio recordings and the corresponding versions captured on common consumer devices such as tablets and smartphones, in real-world environments. We choose the clean-room subset, which includes approximately 4 and 1/2 hours of high-quality data, corresponding to an average of 14 minutes from each of the 20 speakers included in the dataset.
We carefully select 200 segments  (approximately corresponding to utterances) from both male and female speakers, ensuring that each segment is separated by periods of silence.
To ensure compatibility with the D/A converter, we resample the original recordings to 48 kHz.
For the VCTK dataset, we utilize the 96 kHz version\footnote{available at https://datashare.ed.ac.uk/handle/10283/2774} and choose 200 utterances from a diverse group of 24 speakers.

\begin{figure}[!thb]
\centering
\includegraphics[width=0.99\linewidth]{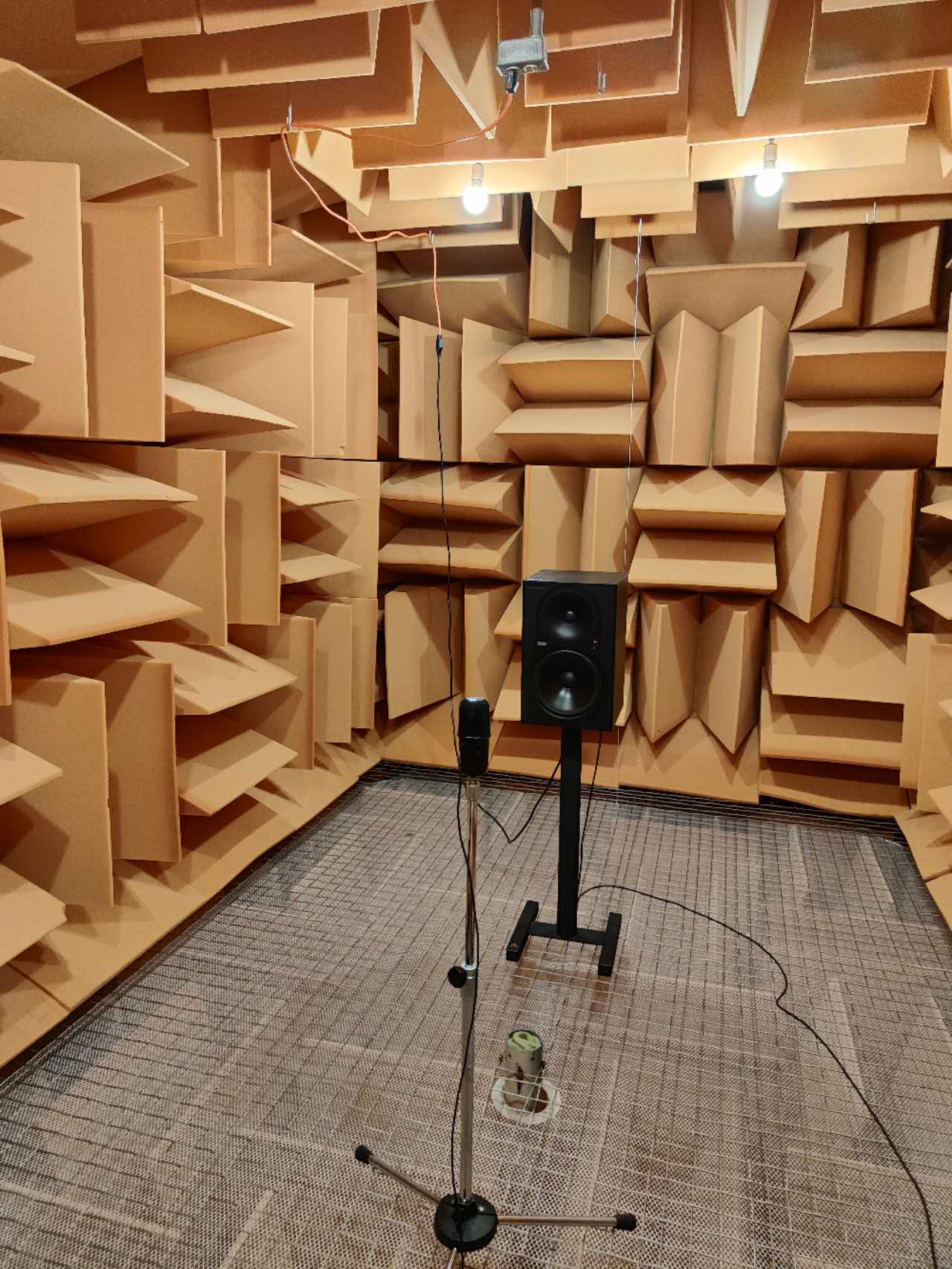}
\caption{Photo of the anechoic chamber and recording setup used for real SR data acquisition.}
\label{fig:room}
\end{figure}

The recordings took place in the anechoic chamber of The Ohio State University Department of Speech and Hearing Science, with the dimensions of 13 x 7.5 x 8 feet (length x width x height), measured from wedge tips. The depth of wedges is 2 feet.
Fig. \ref{fig:room} shows the recording setup and the anechoic chamber.
We conducted three separate recording sessions at the three lowest sampling rates of 8, 16, and 32 kHz of a HyperX SoloCast microphone.
During the recording, we played back the speech utterances from a Windows PC.
The D/A converter employed was an M-Audio Mobile Pre DAC operating at a 48 kHz sampling rate with 24-bit precision. The analog signal was then delivered to a Mackie HR824 powered loudspeaker, renowned for its ``ruler-flat" frequency response (+/- 1.5 dB, 37 Hz – 20k Hz).
The speech signal maintained an approximate level of 65 dBA. We recorded speech signals using the HyperX SoloCast microphone placed 1 meter directly in front of the loudspeaker (see Fig. \ref{fig:room}). The microphone featured onboard A/D conversion, allowing us to capture the digital signal using another Windows PC.
Inside the anechoic chamber, only the loudspeaker and recording microphone were present. All other equipment was located outside the chamber. We had remote control of the apparatus, enabling us to carry out all three recording sessions without re-entering the chamber. Throughout the recording sessions, the lights and fans in the chamber remained off to minimize any external interference.
The recorded datasets are publicly accessible at \href{https://web.cse.ohio-state.edu/~wang.77/pnl/corpus/Heming/RecordedSR.html}{https://web.cse.ohio-state.edu/\(\sim\)wang.77/pnl/corpus/Heming/RecordedSR.html.}

\subsection{Evaluation Metrics}
To measure the SR performance, we use three objective metrics: scale-invariant signal-to-noise ratio (SISNR), log-spectral distance (LSD) \cite{gray1976lsd}, and PESQ for wideband speech \cite{beerends2002perceptual}. SISNR is a time-domain metric that measures signal power relative to noise power in dB.
% SISNR is a time-domain metric that compares the signal power and the noise power in decibels, as follows:
% \begin{equation} 
% \mathrm{SISNR}(s, \hat{s}) = 10 \log_{10} \frac{ \sum_{n=1}^N s(n)^2}{ \sum_{n=1}^N [\hat{s}(n) - s(n)]^2}.
% \end{equation}
LSD is a frequency-domain metric that calculates the logarithmic distance between two magnitude spectra in dB, as follows: 
\begin{equation}
\mathrm{LSD}(S, \hat{S}) = \frac{1}{T} \sum_{t=1}^{T} \sqrt{\frac{1}{F}\sum_{f=1}^F[\log_{10} \frac{|S(t, f)|^2} {|\hat{S}(t, f)|^2}]^2}
\end{equation}
LSD will be 0 dB when two spectra are identical, which is the minimum possible distance.
% PESQ for wideband speech is a standard metric of perceptual speech quality, and its values range from 1.04 to 4.64. Higher PESQ means better listening quality.
PESQ for wideband speech is a standard metric of perceptual speech quality, and higher PESQ means better listening quality.

\subsection{Training Setup}
We use an Adam optimizer \cite{kingma2014adam} to train our model with a batch size of 32 utterances for 100 epochs.
The initial learning rate is 0.0006, and is halved if the validation loss does not improve for three consecutive epochs. Gradient clipping is employed with a maximum value of 1.0 to prevent gradient explosion.
We also apply the exponential moving average to stabilize the training process. 
Within each batch, 4 seconds of each utterance are randomly selected, and shorter utterances are padded with zeros to ensure that all input features are of the same size. Zero-padded parts are discarded in loss calculations. All models are trained on two NVIDIA Volta V100 32GB GPUs, and the DataParallel module from PyTorch \cite{paszke2019pytorch} is used to evenly distribute the batch to the two GPUs during each training step.
In diffusion training, we configure noise scheduling with $\sigma_{min} = 0.05$, $\sigma_{max} = 0.5$, and $\gamma = 1.5$ following the setup in \cite{richter2022sgmse}. 
Training involves 1,000 steps, and inference is capped at 10 steps.

% TODO: address SR-AECNN it is our previous work

\section{Evaluations and Comparisons}
\label{sec:results}
\subsection{Evaluation Results and Comparisons on VCTK}
Table \ref{tbl:baselines} reports the SR performance of the proposed model from 8 kHz to 16 kHz on the VCTK test set, as well as those of four comparison baselines.
The first baseline is a signal processing method that upsamples LR signals to the desired sampling rate using cubic spline interpolation \cite{mckinley1998cubic}.
% To guarantee a fair comparison, we unify the filter designs in testing data simulations by adopting the Chebyshev type I filter with order 8, and adopt the designs that target the 16 kHz sampling rate.
We also compare with three strong DNN-based SR methods, which include a predictive learning based model (our previous study, denoted as SR-AECNN \cite{hm2021towards}), a diffusion-based generative model (NuWave2 \cite{han2022nuwave2}) and a vocoder based model (NVSR \cite{liu2022neural}).
% Proposed better
As shown in the table, our approach achieves the best SISNR of 21.15 dB, and PESQ of 4.05. Its LSD performance also surpasses those of SR-AECNN and NuWave2, demonstrating the advantages of combined predictive and generative modeling.
% Metrics
In addition, all DNN-based approaches show significant LSD improvement over the signal processing baseline.
Although LSD is commonly used for SR evaluations, it does not strongly correlate with the perceptual quality of SR speech, and its use should be complemented by other speech quality metrics such as PESQ.
% For example, in terms of SISNR and PESQ, we observe some degradations for models like NuWave2 and NVSR.
For NuWave2, its SISNR is 2.93 dB worse than the signal processing baseline, and its PESQ drops from 3.44 to 2.71.
% Furthermore, NVSR performs the best in LSD but has poor SISNR and PESQ, this results from adversarial training prioritizing plausible speech components over ground truth, and potential misalignment with the ground truth in vocoder-synthesized speech. 
NVSR obtains superior performance in LSD, but the worst PESQ results. This outcome may be attributed to the misalignment of vocoder-synthesized speech and ground truth signal.
In summary, the proposed model achieves the overall best SR performance.
\begin{table}[!tp]
\centering
\caption{Speech SR performance from 8 to 16 kHz for proposed and baseline methods on the VCTK dataset}
\centering
% \resizebox{0.99\linewidth}{!}{
{
\begin{tabular}{@{}lccc@{}}
\toprule
                 & SISNR & LSD & PESQ \\ \midrule
Cubic Upsampling & 18.99 & 2.72 & 3.44   \\
SR-AECNN            & 20.18 & 0.88 & 3.72  \\
NuWave2          & 19.17 & 1.17 & 2.53  \\
NVSR             & 16.06 & \textbf{0.78} & 2.71  \\
% Proposed         & \textbf{21.29} & 0.80 & \textbf{3.86} \\
Proposed         & \textbf{21.15} & 0.81 & \textbf{4.05} \\

\bottomrule
\end{tabular}
% }
}
\label{tbl:baselines}
\end{table}

% The proposed model supports multiple upsampling ratios by integrating lossmap information.
% To evaluate this, we train a single model on the VCTK dataset and augment the data by simulating LR speech with different upsampling ratios during the training process.
% We also compare it with the other two baselines that support multi-resolution SR, as illustrated in Table \ref{tbl:multi-sr}.
% We display the results of upsampling speech to the target sampling rate 16 kHz with different upsampling ratios of 2, 4 and 8.
% The proposed approach demonstrates a consistent enhancement in terms of LSD in contrast to the cubic spline baseline, while reaching a decent LSD score of 1.44 dB at an upsampling ratio of 8.
% In comparison to NuWave2 \cite{han2022nuwave2}, our model achieves competitive LSD scores but falls slightly short of NVSR \cite{liu2022neural}. 
% However, it is worthwhile to point out that, our proposed approach can achieve stable improvement in SISNR and PESQ values where the other two baselines fail.

% \begin{table*}[!tp]
% \centering
% \caption{LSD Scores for Multi-resolution SR Targeting the 16 kHz Speech, with Upsampling Ratios of 2, 4 and 8}
% \centering
% % \resizebox{0.99\linewidth}{!}{
% {
% \begin{tabular}{@{}lcccc@{}}

% \toprule
%      & 2 & 4 & 8 \\ \cmidrule{2-4}
% Cubic Upsampling & 2.72 & 2.95 & 3.26  \\ 
% \midrule
% NuWave2  & 1.17 & 1.67 & 1.86     \\ 
% NVSR     & 0.78 & 0.95 & 1.07 \\ 
% % *Proposed & 0.85 & 1.23 & 1.64 \\ DDPM
% Proposed & 0.83 & 1.20 & 1.44 \\  % DDRM

% \bottomrule
% \end{tabular}
% % }
% }
% \label{tbl:multi-sr}
% \end{table*}

To further understand the effects of two-stage training, we conduct an ablation study to compare the two-stage model to single-stage models on the SR task.
The findings of this analysis are presented in Table \ref{tbl:ablation}. 
Two single-stage models correspond to predictive learning only with DPARN and diffusion-based learning only with ARCN.
The ablation results indicate that the predictive learning model outperforms the generative model, especially in SISNR and PESQ.
The proposed two-stage approach yields consistently better results than the single-stage models.

\begin{table}[!tp]
\centering
\caption{Single-stage versus two-stage models}
\centering
% \resizebox{0.99\linewidth}{!}{
{
\begin{tabular}{@{}lccc@{}}
\toprule
                  & SISNR & LSD & PESQ \\ \midrule
Cubic Upsampling  & 18.99 & 2.72 & 3.44  \\
DPARN             & 20.89 & 0.82 & 3.87  \\
ARCN              & 19.12 & 0.85 & 3.71  \\
% DPARN  + ARCN     & 21.29 & 0.80 & 3.86  \\ % DDPM
DPARN + ARCN     & 21.15 & 0.81 & 4.05  \\ % DDRM

\bottomrule
\end{tabular}
% }
}
\label{tbl:ablation}
\end{table}

\subsection{Robustness Evaluation}

\begin{table*}[!tp]
\centering
\caption{Evaluation on LR speech generated by different filters}
\centering
{
\begin{tabular}{@{}lccccccc@{}}

\toprule
           & \multicolumn{3}{c}{Chebyshev} && \multicolumn{3}{c}{Bessel}\\ \cmidrule{2-4} \cmidrule{6-8}
             
                 & SISNR & LSD & PESQ     && SISNR & LSD & PESQ \\ \midrule
Cubic Upsampling &  18.99 & 2.72 & 3.44    && 17.21 &  2.89 & 3.33   \\ 
\midrule
SR-AECNN              & 20.18 & 0.88 & 3.72    && -8.84 &  2.43 & 1.07 \\ 
NuWave2            & 19.17 & 1.17 & 2.53    && 10.83 &  1.86 & 1.77 \\ 
NVSR               & 16.06 & 0.78 & 2.71    && 13.64 &  1.05 & 2.77 \\ 
% Proposed         & 21.29 & 0.80 & 3.86    && 17.14 &  1.25 & 3.47  \\ %DDPM
Proposed         & 21.15 & 0.81 & 4.05    && 17.43 &  1.23 & 3.52  \\ %DDRM
\bottomrule
\end{tabular}
}
% }
\label{tbl:filter}
\end{table*}

Our previous study \cite{hm2021towards} reveals that predictive learning DNNs are sensitive to specific downsampling filters used in training.
Now, we assess the filter robustness of the proposed model and other baselines for the SR task from 8 kHz to 16 kHz on the VCTK test set.
Specifically, we train the models on Chebyshev filter-simulated data and test them on LR speech generated by a Chebyshev filter (matched) and a fifth-order Bessel filter (mismatched).
As shown in Table \ref{tbl:filter}, our proposed model exhibits good performance in both matched and mismatched conditions.
Compared to cubic spline upsampling, our model demonstrates consistent improvements across SISNR, LSD, and PESQ metrics.
SR-AECNN shows severe performance degradation in the mismatched condition.
% Add results of NuWave and NVSR
NuWave2 also shows a considerable performance drop.
NVSR, on the other hand, has a relatively stable performance across the two conditions as it leverages a vocoder to synthesize the SR speech. It has the best LSD scores, but its SISNR and PESQ scores are even worse than the cubic spline baseline.

\subsection{Evaluation Results and Comparisons on Recorded Data}

\begin{table*}[!tp]
\centering
\caption{Speech SR performance from 8 to 16 kHz and from 8 to 32 kHz on recorded data }
\centering
% \resizebox{0.99\linewidth}{!}{
{
\begin{tabular}{@{}lccccccccccccccc@{}}

\toprule
     & \multicolumn{7}{c}{8 kHz $\rightarrow$ 16 kHz} && \multicolumn{7}{c}{8 kHz $\rightarrow$ 32 kHz} \\ \cmidrule{2-8} \cmidrule{10-16}
     & \multicolumn{3}{c}{DAPS} && \multicolumn{3}{c}{VCTK} && \multicolumn{3}{c}{DAPS} && \multicolumn{3}{c}{VCTK} \\ \cmidrule{2-4} \cmidrule{6-8} \cmidrule{10-12} \cmidrule{14-16} 
             
                 & SISNR & LSD & PESQ       && SISNR & LSD & PESQ   && SISNR & LSD & PESQ && SISNR & LSD & PESQ\\ \midrule
Cubic Upsampling & 12.68  & 3.34 & 3.23     && 12.53 & 3.37 &  3.06  && 12.40 & 4.39 &  2.62 && 10.97 & 4.93 &  2.33      \\ 
\midrule
SR-AECNN            & -10.51 & 2.09 & 1.29  && 10.45   & 1.37  & 1.98 && -5.06   & 1.41 & 1.32 && 6.61  & 1.50 & 1.52  \\ 
NuWave2          & 12.17  & 1.17 & 2.84  && 11.89   & 1.06  & 2.41 && 10.43   & 1.38  & 1.45 && 8.83    &  1.48  & 1.35    \\ 
NVSR             & 10.87  & 1.15 & 1.92  && 10.79   & 1.07  & 2.02 && 10.69   & 1.15  & 2.04 && 8.98   & 1.06  & 2.08     \\ 
% Proposed    & 12.42  & 1.08 & 3.64  && 13.91   & 1.00  & 3.61 \\ % DDPM
Proposed         & 12.34  & 1.09 & 3.68  && 13.97  & 1.02  & 3.64  && 12.45   & 1.19  & 3.07 && 11.50  & 1.23  & 2.93 \\ % DDRM

\bottomrule
\end{tabular}
% }
}
\label{tbl:real-sr}
\end{table*}

Finally, we evaluate on the two recorded datasets described in Section \ref{sec:real_dataset}.
% The gold standard to assess real-world applicability of DNN models is the evaluation of real recordings.
We present the results in Table \ref{tbl:real-sr} for two SR tasks: from 8 kHz to 16 kHz,  and from 8 kHz to 32 kHz.
All the models are trained on the VCTK dataset with the corresponding upsampling ratios.
Comparing to the results in Table \ref{tbl:baselines}, all SR methods show reduced performance on the recorded VCTK for the task of 8 kHz $\rightarrow$ 16 kHz, reflecting the mismatch between simulated and recorded data. 
For both SR tasks, our model shows a consistent improvement compared to the cubic upsampling method. In addition, the proposed model outperforms other baselines in all metrics except for NVSR which produces better LSD scores for the 8 kHz $\rightarrow$ 32 kHz task.
The SR-AECNN model yields large performance deterioration, particularly on the DAPS dataset.
In contrast, generative and vocoder based models deliver relatively stable SR performance.
It is worth noting that, while NuWave2 and NVSR produce strong LSD scores, their SISNR and PESQ scores are not only substantially worse than our model but also than the signal processing baseline.
This evaluation suggests that our proposed approach overcomes the limitation of training with simulated data, and can be applied in real-world environments.

\section{Conclusion}
\label{sec:conclusion}

In this study, we have introduced a two-stage model that combines predictive and generative approaches for speech SR.
Our model consistently outperforms other strong baselines in both simulated and realistic datasets.
The model is capable of generating high-frequency speech in mismatched conditions, making it suitable for real-world applications. 
An additional contribution is the collection and publication of recorded data using the same microphone at multiple native sampling rates, to facilitate SR progress in real-world environments.

% or


% \appendices

% % you can choose not to have a title for an appendix
% % if you want by leaving the argument blank
% \section{}

\ifCLASSOPTIONcompsoc
  \section*{Acknowledgments}
\else
  \section*{Acknowledgment}
\fi
% This research was supported in part by an NIDCD grant (R01 DC012048) and the Ohio Supercomputer Center.
This research was supported in part by an NIDCD (R01 DC012048) grant, the Ohio Supercomputer Center, and the Pittsburgh Supercomputing Center (under NSF grant ACI-1928147).
\ifCLASSOPTIONcaptionsoff
  \newpage
\fi



% http://www.michaelshell.org/tex/bibtex/
%\bibliographystyle{IEEEtran}


% \end{thebibliography}

\bibliographystyle{IEEEtranS}
\bibliography{refs}

\end{document}